\documentclass[sigplan, screen]{acmart}

\copyrightyear{2027}
\acmYear{2027}
\setcopyright{cc}
\setcctype{by}
\acmConference[ASPLOS '27]{Proceedings of the 32nd ACM International Conference on Architectural Support for Programming Languages and Operating Systems, Volume 1}{April 11--15, 2027}{Heraklion, Greece}
\acmBooktitle{Proceedings of the 32nd ACM International Conference on Architectural Support for Programming Languages and Operating Systems, Volume 1 (ASPLOS '27), April 11--15, 2027, Heraklion, Greece}
\acmDOI{10.1145/3845814.3850072}
\acmISBN{979-8-4007-3012-2/2027/04}

\usepackage{tikz}
\usepackage{amsmath}

\usepackage{booktabs}   
\usepackage{subcaption} 

\usepackage{microtype}
\usepackage{enumitem}
\usepackage{algorithmic}
\usepackage[ruled,linesnumbered, vlined]{algorithm2e}
\usepackage{multirow}
\usepackage[table]{xcolor}
\usepackage{pifont}
\usepackage{bm}
\usepackage{balance}
\usepackage{xspace}
\usepackage{xurl}

\newcommand{\revised}[1]{\textcolor{black}{{}#1}} 
\newcommand{\reviseded}[1]{\textcolor{black}{{}#1}} 
\newcommand{\sysname}{\textsc{EdgeAgent}}
\date{}

\author{Yuhai Long}
\email{longyh27@mail2.sysu.edu.cn}
\affiliation{%
  \institution{Sun Yat-sen University}
  \department{School of Computer Science and Engineering}
  \city{Guangzhou}
  \country{China}
}

\author{Yuanxin Wei}
\email{weiyx25@mail2.sysu.edu.cn}
\affiliation{%
  \institution{Sun Yat-sen University}
  \department{School of Computer Science and Engineering}
  \city{Guangzhou}
  \country{China}
}

\author{Kai Wu}
\email{wukai6@cmic.chinamobile.com}
\affiliation{%
  \institution{China Mobile Internet Company Ltd.}
  \city{Guangzhou}
  \country{China}
}

\author{Jinhui Wei}
\email{weijh28@mail2.sysu.edu.cn}
\affiliation{%
  \institution{Sun Yat-sen University}
  \department{School of Computer Science and Engineering}
  \city{Guangzhou}
  \country{China}
}

\author{Dan Huang}
\email{huangd79@mail.sysu.edu.cn}
\affiliation{%
  \institution{Sun Yat-sen University}
  \department{School of Computer Science and Engineering}
  \city{Guangzhou}
  \country{China}
}

\author{Jiangsu Du}
\email{dujiangsu@mail.sysu.edu.cn}
\authornote{Corresponding author.}
\affiliation{%
  \institution{Sun Yat-sen University}
  \department{School of Computer Science and Engineering}
  \city{Guangzhou}
  \country{China}
}

\begin{document}

\title[EdgeAgent: On-Device LLM inference for Multi-Agent Systems]{EdgeAgent: Orchestrating On-Device LLM inference for End-User Multi-Agent Systems on CPU-GPU Unified Memory Architectures}

\begin{abstract}
Emerging multi-agent LLMs demand privacy-preserving edge deployment, yet current inference systems struggle with these collaborative workflows. Specifically, the memory-bound decode phase causes severe bus contention on unified memory architectures (UMA), paralyzing naive CPU-GPU co-execution. Furthermore, speculative decoding in multi-agent workloads faces extreme variance in drafting difficulty, alternating between complex reasoning and predictable structured generation. Compounded by frequent tool-induced stalls, this highly fragmented execution severely underutilizes hardware and defeats traditional static batching.

We present \sysname{}, a cross-layer inference system explicitly co-designed for edge UMA and multi-agent workloads. At the micro-architectural level, it bypasses rigid graph-compiler constraints to enable zero-copy UMA-aware tensor parallelism, utilizing asymmetric memory layouts to fully saturate both CPU and GPU compute units. At the scheduling level, it dynamically allocates draft budgets based on real-time sequence predictability to bound bandwidth waste. Concurrently, an asynchronous suspend-and-yield mechanism actively evicts stalled agents, ensuring continuous hardware saturation during unpredictable tool invocations.

\revised{Extensive evaluations on an Apple M4 SoC demonstrate that the UMA-aware execution alone contributes a 1.29× speedup over batched speculative decoding. Adding the agent-aware scheduling lifts the full EdgeAgent system to a 1.77× speedup under extreme tool-use latencies.}

\end{abstract}


          
\maketitle

\section{Introduction}

The application of large language models~\cite{brown2020language, deepseek2024v2, gerganov2023llama} (LLMs) is increasingly shifting from monolithic chat interfaces toward coordinated multi-agent systems.
In recent agent frameworks, such as LangGraph~\cite{langchain2026github} and OpenClaw~\cite{openclaw2026github}, a common design choice is to decompose complex tasks into specialized roles and coordinate them through explicit orchestration mechanisms.
With the enhanced capability of multi-agent systems, they start to take over more system access privileges, raising growing privacy and security concerns.
This trend makes on-device LLM inference deployment increasingly important.
However, existing LLM inference systems can hardly handle such multi-agent workloads on end-user devices efficiently.

Most existing LLM inference systems~\cite{kwon2023efficient, qin2025mooncake, yu2022orca} are primarily designed for cloud environments and multi-user workloads. These systems process hundreds of concurrent requests across multiple discrete GPUs, targeting high-through\-put batching and multi-tenant isolation. 
Lacking awareness of fine-grained, intra-node memory contention, their hardware and workload assumptions render them highly inefficient when ported directly to edge devices.
Conversely, existing edge frameworks (e.g., Llama.cpp~\cite{gerganov2023llama}, MLX~\cite{apple2023mlx}) primarily target single-user, single-request sequential inference. 
When confronted with collaborative multi-agent workloads which dynamically generate concurrent, highly fragmented requests on a single device, these monolithic edge systems lack the necessary orchestration, leaving available compute resources severely underutilized.


The more balanced CPU-GPU computing capability on modern edge devices enables effective use of tensor parallelism (TP) to coordinate heterogeneous execution and fully utilize on-chip compute resources~\cite{jiang2025doppeladler, wei2025ghidorah, li2025accelerating, chen2025characterizing, wei2025agentxpu}.
Modern edge systems-on-chip (SoCs) typically integrate powerful mobile GPUs and highly capable CPUs with advanced matrix extensions, such as ARMv9 Scalable Matrix Extension~\cite{arm2021sme} (SME) in the Apple M4~\cite{apple2024m4}, forming a well-proportioned heterogeneous computing platform. 
TP~\cite{narayanan2021efficient}, as a mainstream model parallelism technique, partitions LLM workloads (e.g., matrix multiplications) across multiple devices, making it suitable for distributing computation between CPUs and GPUs to speed up inference. 
During the compute-bound prefill phase, heterogeneous TP can fully exploit aggregated floating-point throughput by engaging both processing units simultaneously. 
However, during the decode phase, such CPU-GPU parallelism brings little performance gain under the shared unified memory architecture (UMA), since the decoding process is inherently memory-intensive and severely bounded by memory bandwidth.


To mitigate the memory bound in the decode phase, recent systems improve arithmetic intensity through speculative decoding (SD)~\cite{li2025eagle3, cai2024medusa}. SD decouples generation into two stages: a lightweight draft model first proposes multiple candidate tokens, and the target model then verifies them in parallel, accepting the longest matching prefix. By amortizing target-model weight loading across multiple token verifications, SD reduces per-token bandwidth pressure and improves decode throughput.

However, integrating these hardware-aware optimizations for multi-agent workflows into existing graph-based edge frameworks, such as MLX, exposes three critical systemic bottlenecks.
First, implementing CPU-GPU lock-free TP on shared memory exposes the limitations of modern graph compilers~\cite{lattner2021mlir, ansel2024pytorch2}. Standard directed acyclic graph (DAG) executors cannot safely express asynchronous, orthogonal memory mutations to a single shared tensor, forcing runtimes into eager synchronization which severely disrupts the graph optimization pipeline and explicit memory copies that negate the zero-copy benefits of UMA.
Second, different agent tasks exhibit extreme variance in the difficulty of drafting (e.g., highly structured code synthesis vs. deliberate reasoning). 
As this drafting difficulty is related to the real-time context, traditional SD systems allocate a static draft length uniformly across all tasks. This inevitably results in the waste of limited memory bandwidth, which is frequently used to process rejected drafts.
Third, agents frequently execute external tool calls~\cite{yao2023react, liu2024agentbench} that incur out-of-system stall. When an agent stalls on a tool call, the runtime cannot dynamically redistribute its idle speculative budget to remaining active agents, failing to fully exploit the available compute capacity and degrading overall system performance on edge devices.

To address these challenges, we propose \sysname{}, a co-designed inference system explicitly tailored for edge UMA and concurrent multi-agent workflows. 
\sysname{} adopts a cross-layer architecture featuring an agent-aware scheduling layer, which tightly couples real-time agent states and task predictability directly with the underlying UMA-aware execution layer.
In summary, this paper makes the following contributions:
\begin{itemize}[leftmargin=*]
    \item We design a UMA-aware execution layer with a zero-copy TP graph that adopts symmetric memory layouts to eliminate compiler overhead, together with optimized SME micro-kernels to jointly saturate heterogeneous edge compute units.
    \item We design an agent-aware scheduling layer that dynamically allocates draft budgets based on real-time drafting difficulty, and employs an in-place suspend-and-yield mechanism to instantly reclaim computation slots during tool-induced stalls.
    \item We implement and evaluate \sysname{} on modern edge architectures using trace-driven, concurrent multi-agent workflows. Experimental results demonstrate that our system delivers substantial improvements in both global concurrent throughput and end-to-end task make\-span compared to state-of-the-practice inference frameworks.
\end{itemize}

\section{Background and motivation}
\label{sec:motivation}

\subsection{LLM Inference Process}
\begin{figure}[!t]
    \centering
    \includegraphics[width=0.98\linewidth]{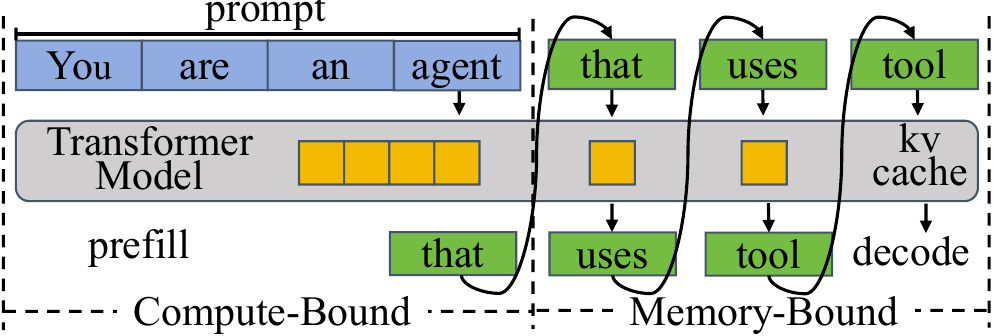}
    \caption{LLM autoregressive decoding process.}
    \label{fig:llminfer}
\end{figure}

As illustrated in Figure~\ref{fig:llminfer}, LLM inference proceeds in an autoregressive manner, generating tokens sequentially until an end-of-sequence (EoS) token is reached. To improve efficiency, the key and value representations from previous steps are cached in memory (i.e., the KV cache), allowing subsequent steps to avoid recomputing them.
Based on this mechanism, the inference process can be divided into two stages: prefill and decode. 
The prefill stage processes the input prompt and constructs the initial KV cache. Since the prompt usually contains multiple tokens, this stage exhibits substantial parallelism. In contrast, during the decode stage, the model generates one token at a time. Each step consumes the existing KV cache, produces a new token, and appends the corresponding KV entries.
In end-user settings, inference is typically performed on a single request, where each decoding step processes only one token. As a result, the decode stage becomes largely memory-bandwidth-bound, limiting hardware utilization. Furthermore, because decoding proceeds token by token, it often constitutes the dominant portion of the overall inference latency.

\subsection{Multi-Agent Inference Workloads}
\label{sec:agent_evolution}

\begin{figure}[!t]
    \centering
    \includegraphics[width=\linewidth]{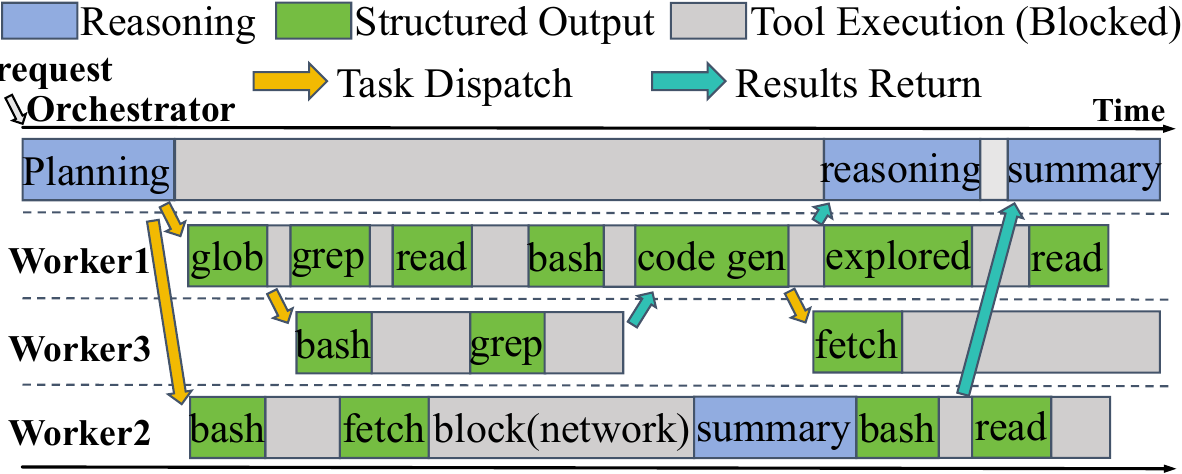}
    \caption{Multi-Agent workload.}
    \label{fig:multi-agent}
\end{figure}

Unlike traditional LLM applications functioning as chatbots for sequential queries, the adoption of LLMs is rapidly shifting to a collaborative, multi-agent system.
This shift starts from ReAct~\cite{yao2023react}, which integrates LLM's internal reasoning with external tools to solve complex tasks.
In such a paradigm, completing a single user request requires coordination among multiple specialized agents, each exhibiting different workload characteristics, as illustrated in Figure~\ref{fig:multi-agent}.
For instance, an \textit{Orchestrator} agent serves as the central planner, processing extensive historical contexts to formulate strategies via complex reasoning. As these open-ended reasoning traces lack rigid structures, they are difficult to draft speculatively. 
In contrast, \textit{Worker} agents handle concrete execution by translating these plans into structured executable code or rigid tool-use instructions. Driven by fixed syntactic templates, their outputs are highly structured and consequently much easier to draft.

Moreover, as Worker agents must wait for responses from external tools (e.g., executing scripts or querying APIs), they incur severe execution stalls. 
Ultimately, the multi-agent workload exposes the inference engine to a highly fragmented execution mode, characterized by significant differences in drafting difficulty and unpredictable temporal blocking. This fundamentally differs from the assumption of continuous, homogeneous processing in traditional LLM serving~\cite{yu2022orca, kwon2023efficient}, posing challenges for efficient execution of inference systems.

\subsection{Speculative Decoding}
\label{sec:speculative_decoding}

\begin{figure}[!t]
    \centering
    \includegraphics[width=0.98\linewidth]{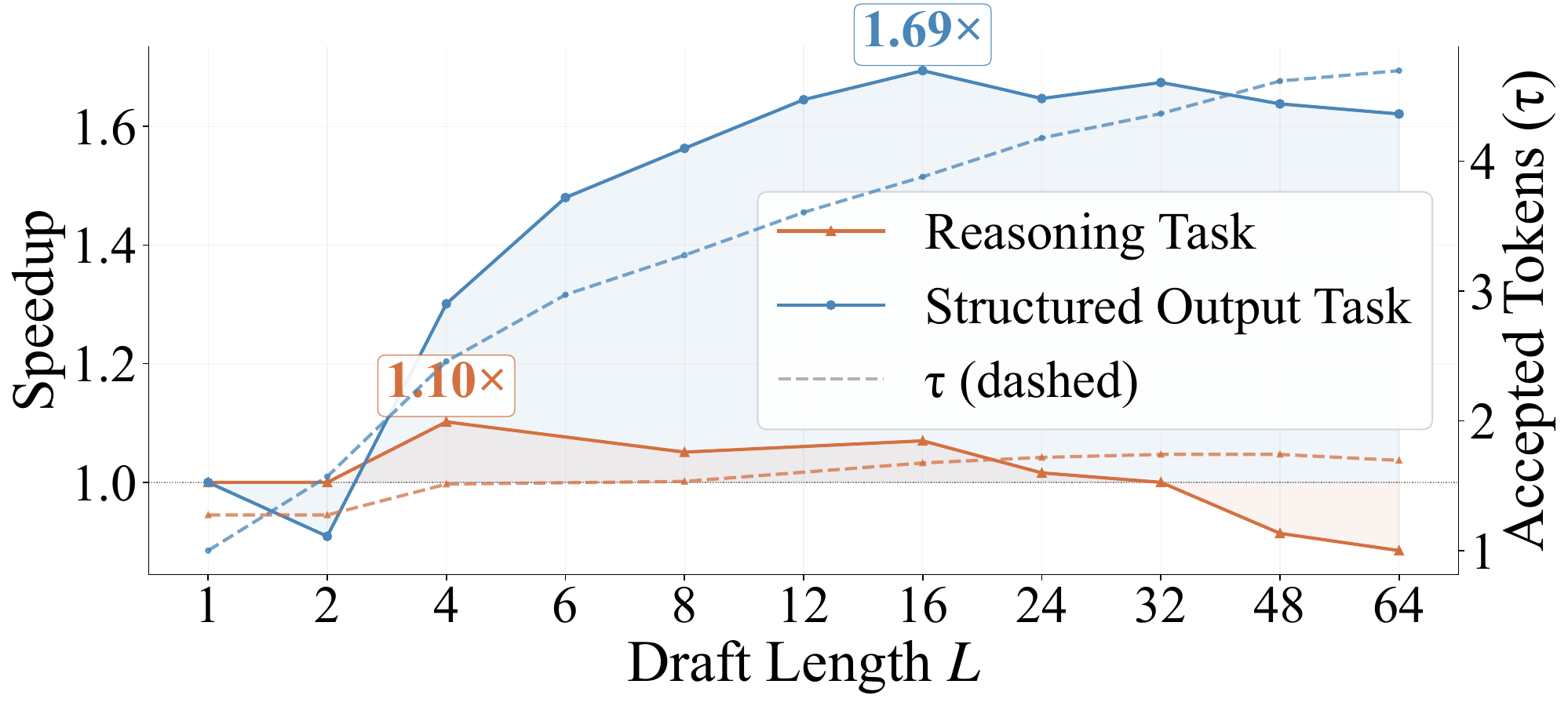}
    \caption{Task-dependent non-linear speedup of speculative decoding on edge UMA: accepted tokens ($\tau$) increase sub-linearly with draft length $L$, while the end-to-end speedup increases first and then degrades due to rejection overhead.}
    \label{fig:spec_nonlinear}
\end{figure}


To mitigate the memory bandwidth bottleneck of autoregressive generation, speculative decoding (SD) decouples inference into a lightweight drafting phase and a parallel verification phase. In this paradigm, a small draft model predicts candidate token sequences that the massive target model subsequently verifies in a single forward pass. This design effectively amortizes the high memory latency associated with loading target model weights from main memory, as the memory-intensive target model is invoked only once per draft batch rather than sequentially for each token.

However, the speedup achieved by SD scales non-linearly with draft length $L$, and this pattern varies significantly across different tasks.
As illustrated in Figure~\ref{fig:spec_nonlinear}, we conduct a pre-experiment on an Apple M4 SoC, deploying the DeepSeek-R1-Distill-Llama-8B model. We evaluate the acceleration profiles across two highly contrasting multi-agent workloads: deliberate reasoning (sampled from LongBench~\cite{bai2025longbench}) and highly predictable structured generation (sampled from MBPP~\cite{austin2021program}).
Empirically, the speedup initially surges as memory traffic is amortized across more accepted tokens.
The accepted token rate $\tau$ increases sub-linearly with $L$, and the end-to-end speedup peaks at $L = 16$ with a maximum speedup of $1.69\times$ on the Structured Output Task.
Beyond this optimal length, the speedup diminishes rapidly and even degrades, driven by the surging probability of consecutive draft rejections. 
Critically, this non-linear behavior is task-dependent.
Tasks with low drafting difficulty (e.g., structured output generation) tolerate longer draft lengths and achieve higher peak speedups, while hard reasoning tasks experience diminishing returns at shorter $L$ due to frequent token rejections. 
This implies that different sub-agents require distinct optimal draft lengths to maximize acceleration, as their drafting difficulties and token predictability vary significantly.

Unfortunately, existing SD frameworks enforce a uniform draft length across all requests within a batch to simplify kernel execution, which further leads to resource misallocation. 
Driven by the significant differences in drafting difficulty across various multi-agent tasks (\S\ref{sec:agent_evolution}), this one-size-fits-all budget allocation brings two critical issues.
On one hand, generating deep drafts for hard-to-draft reasoning tasks wastes shared memory bandwidth, as many candidate tokens are ultimately rejected. 
On the other hand, batch-level constraints truncate highly predictable tool instruction generation, limiting their maximum theoretical token-acceptance rates.


\subsection{The Edge UMA Memory Wall}
\label{sec:memory_wall}

Modern SoCs (e.g., Apple M4, Snapdragon X~\cite{qualcomm2023snapdragonx}) feature unified memory architectures (UMA) and advanced tensor extensions. 
By abandoning discrete topologies, UMA allows the CPU and GPU to share a single physical memory pool and bus interface. Simultaneously, equipped with extensions like ARMv9 SME, modern CPUs now rival mobile GPUs in dense computation, even outperforming them at small batch sizes. 
Though this convergence supports zero-copy sharing of model weights across heterogeneous cores, it triggers memory contention.
Specifically, during the prefill phase, high arithmetic intensity keeps the system compute-bound, allowing heterogeneous tensor parallelism (TP) to effectively aggregate compute power. However, in the decode phase, extremely small batch sizes shift the bottleneck entirely to memory bandwidth, fundamentally degrading the efficiency of hardware concurrency.

\begin{table}[!t]
\caption{Memory wall on Apple M4 UMA for a uniform read payload: CPU--GPU co-run speedup diminishes as the matrix scales.}
\label{tab:bandwidth_probe}
\centering
\small
\setlength{\tabcolsep}{4pt} 
\begin{tabular}{c|ccc|cc}
\toprule
\multirow{2}{*}{\textbf{Matrix ($K \times N$)}} & \multicolumn{3}{c|}{\textbf{Bandwidth (GB/s)}} & \multicolumn{2}{c}{\textbf{Speedup}} \\ \cmidrule(lr){2-4} \cmidrule(l){5-6}
 & \textbf{CPU} & \textbf{GPU} & \textbf{Co-run} & \textbf{vs CPU} & \textbf{vs GPU} \\ \midrule
4096 $\times$ 4096    & 66.1 & 54.3 & 73.9 & 1.12$\times$ & 1.36$\times$ \\ 
4096 $\times$ 14336   & 68.9 & 80.3 & 91.3 & 1.32$\times$ & 1.14$\times$ \\ 
14336 $\times$ 4096   & 68.8 & 80.6 & 90.8 & 1.32$\times$ & 1.13$\times$ \\ 
4096 $\times$ 28672   & 71.5 & 88.9 & 93.4 & 1.31$\times$ & 1.05$\times$ \\ 
4096 $\times$ 128256  & 72.0 & 96.7 & 95.4 & 1.32$\times$ & 0.99$\times$ \\ \bottomrule
\end{tabular}
\end{table}

As in Table~\ref{tab:bandwidth_probe}, empirical micro-benchmarks on the Apple M4 SoC confirm the severe UMA bandwidth constraint. 
A strong scaling test shows that for small matrices, CPUs outperform GPUs and the co-run achieves a speedup of $1.36\times$ over GPU on the $4096 \times 4096$ matrix.
However, as the matrix size scales, GPUs approach the UMA’s effective bandwidth ceiling (utilizing over 80\% of the 120 GB/s peak). 
Therefore, CPU-GPU co-run suffers from destructive bus arbitration, resulting in a speedup of $0.99\times$ over GPU on the $4096 \times 128256$ matrix.
This bandwidth wall explains why heterogeneous TP during the decode phase yields negligible end-to-end speedup, necessitating an agent-aware global co-scheduling paradigm to bypass the bottleneck.

\subsection{Graph Compiler Limitations}
\label{sec:graph_compiler}

Beyond the hardware memory wall, realizing fine-grained device coordination is obstructed by existing software infrastructure. The modern deep learning ecosystem relies heavily on graph compilation as the foundational software for high-performance inference. 
This paradigm manifests through distinct paths across frameworks: PyTorch introduced ExecuTorch~\cite{executorch2023github} for ahead-of-time (AOT) static graph serialization, whereas Apple's native MLX defaults to global lazy evaluation to dynamically construct computation graphs at runtime. Despite these divergent implementations, both approaches transparently construct a global directed acyclic graph (DAG) of the program. Leveraging this intermediate representation, backend compilers can execute aggressive operator fusion and in-place memory lifecycle reuse, drastically reducing the framework overhead associated with per-operator scheduling. 

While graph compilation has achieved tremendous success in optimizing for standalone hardware backends, its underlying DAG abstraction exposes severe system-level limitations for fine-grained, heterogeneous TP across the CPU and GPU on edge devices. The core deficiency of modern graph compilers lies in their coarse, tensor-level dependency tracking, which inherently struggles to safely express the lock-free, concurrent mutation of a single shared physical buffer. Within edge UMA architectures, avoiding exorbitant memory copy overheads requires the CPU and GPU to perform orthogonal writes to different offset regions of a contiguous memory block. However, because the DAG tracks data hazards strictly at a tensor-level granularity, it is semantically blind to these disjoint sub-tensor boundaries and flags these orthogonal writes as severe data conflicts. Unable to semantically guarantee the safety of such non-overlapping mutations, the state machine of the graph executor resorts to a conservative over-protection strategy. It forcibly inserts coherence hardware fences across device boundaries, compelling the local graph execution to fall back into an eager synchronization mode. This semantic gap within the DAG implementation induces severe interleaved blocking among heterogeneous instruction sequences, significantly diminishing the parallel speedups gained from underlying operator partitioning.

\section{\sysname{} Design and Implementation}
\label{sec:design}

\begin{figure}[!t]
    \centering
    \includegraphics[width=0.95\linewidth]{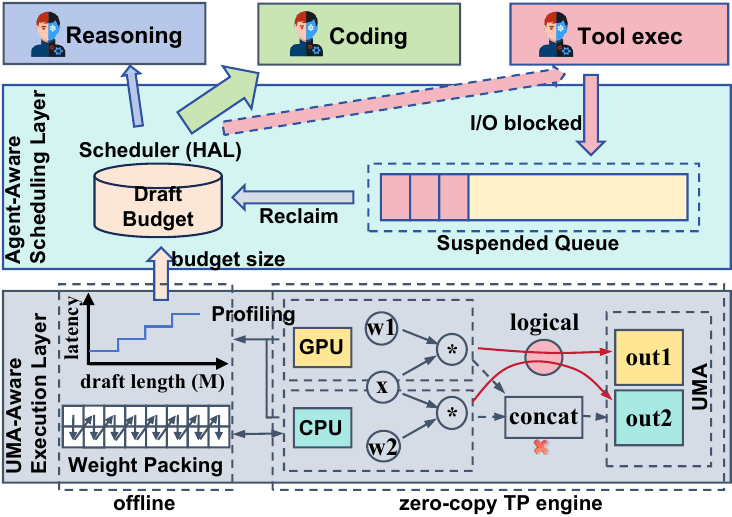}
    \caption{The architecture overview of \sysname{}}
    \label{fig:architecture}
\end{figure}

As shown in Figure \ref{fig:architecture}, \sysname{} acts as a cross-layer inference system that couples the multi-agent workflows with the hardware-level physical constraints of edge UMA.
Instead of processing requests uniformly, \sysname{} disaggregates each request's resource scheduling from overall execution to prioritize resource allocation to demand-critical requests and improve overall resource utilization. The architecture is organized into two collaborative layers, i.e. execution layer and scheduling layer.

The \textit{UMA-aware execution layer} (Section \ref{sec:hetero_tp}) provides a unified heterogeneous backend to maximize hardware saturation via: (1) A zero-copy TP computation graph, which features asymmetric memory layouts for heterogeneous processors, plus custom DAG barriers that bypass compiler overheads to enable lock-free tensor parallelism (Section \ref{sec:zero_copy_TP}). (2) SME micro-kernel optimizations, which combine instruction-level pipelining, cache-aware tiling, and dynamic work-stealing to maximize CPU efficiency (Section \ref{sec:sme}) .

The \textit{agent-aware scheduling layer} (Section \ref{sec:scheduling_layer}) employs a dynamic global scheduler to maintain optimal hardware saturation via two mechanisms.
First, to maximize throughput with limited draft slots, it dynamically allocates budgets toward highly predictable sequences using the Historical Accepted Length (HAL) as an indicator of drafting difficulty (Section \ref{sec:dynamic_spec}).
Second, to mitigate tool-induced stalls, \sysname{} employs a cooperative suspend-and-yield mechanism to detect external tool calls and reclaim idle compute capacity for other active requests.
By promptly decoupling stalled agents from the active batch pool and backfilling their execution slots with pending tasks, the scheduler keeps the hardware well utilized (Section \ref{sec:suspend_yield}).

\subsection{UMA-Aware Execution Layer}
\label{sec:hetero_tp}

\begin{figure}[!t]
    \centering
    \includegraphics[width=0.95\linewidth]{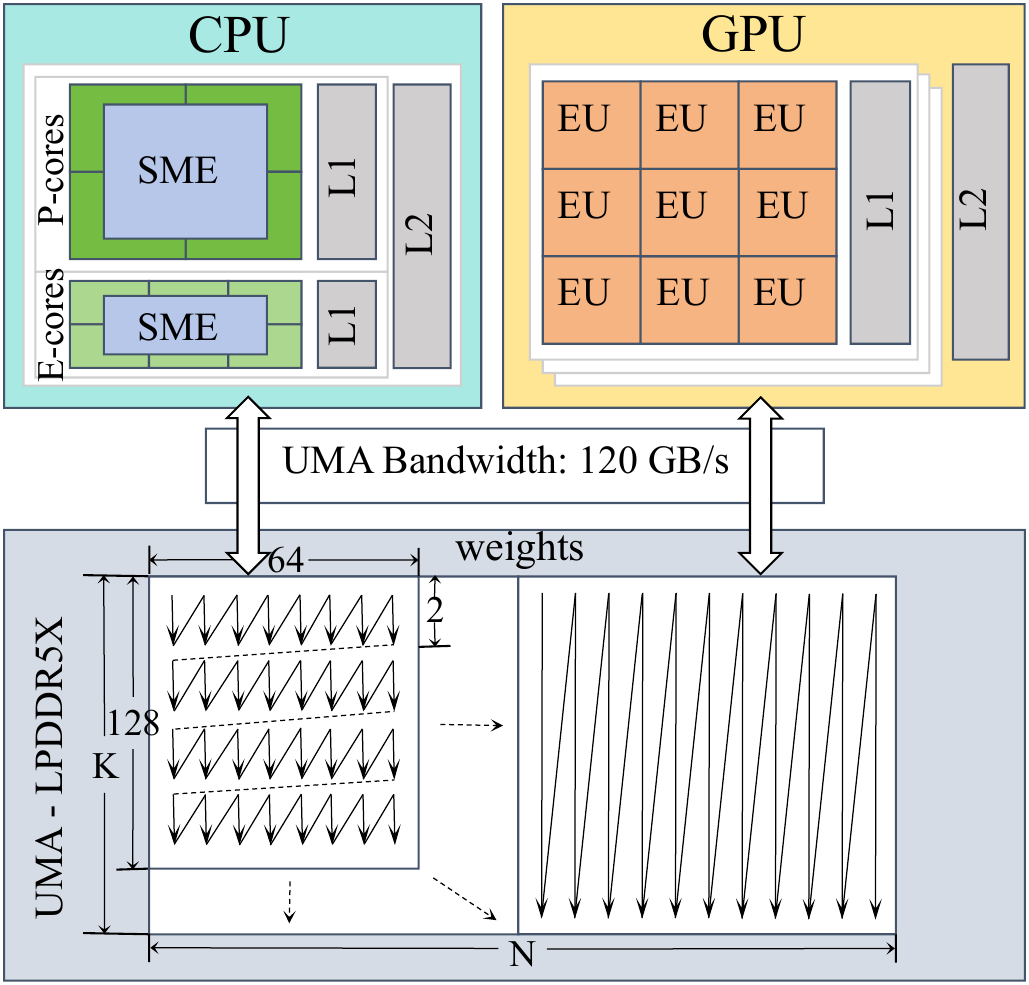}
    \caption{Asymmetric weight layouts in \sysname{} for zero-copy heterogeneous TP on the Apple M4 SoC. CPU shards are pre-packed into 16×64 FP16 tiles to maximize SME utilization, while GPU shards retain a linear SIMT layout.}
    \label{fig:uma_and_rearrange_weight}
\end{figure}

\subsubsection{Zero-Copy TP}
\label{sec:zero_copy_TP}

We design a UMA-aware heterogeneous computation graph that maximizes TP performance on edge devices through two key structural optimizations. First, to bridge the micro-architectural gap between processors, the graph enforces an asymmetric weight layout (Figure~\ref{fig:uma_and_rearrange_weight}). GPU shards retain a standard linear layout for SIMT, while CPU shards are offline pre-packed into 2D tiles for SME, satisfying the memory stride constraints of both units. Second, to eliminate memory copy overhead, we allow both devices to concurrently write their partial output results directly into a single shared physical buffer.

\revised{\paragraph{Model partitioning and weight residency.}
\sysname{} partitions each transformer block with two patterns, both fixed at load time. (1) Every linear projection, including the MLP gate/up/down projections and the attention Q/K/V/O projections, is column-parallel along the output (N) dimension. A weight matrix $W \in \mathbb{R}^{K\times N}$ is split along its N (output-column) dimension into $W_{cpu}\in\mathbb{R}^{K\times n}$ and $W_{gpu}\in\mathbb{R}^{K\times(N-n)}$,
where $n=\alpha\cdot N$ and each layer's split ratio $\alpha$ is computed from its K/N
dimensions to equalize CPU and GPU compute time. Both devices read the same activation
from shared memory and write disjoint column ranges of one preallocated output buffer,
so no cross-device reduction is ever needed. (2) The attention core is partitioned by
heads: the GPU runs a fused SDPA over its head range while the CPU runs a grouped matmul
over the remaining heads, with the per-head score/softmax/context computation entirely
device-local. The two head-group contexts are then recombined by a single concat along
the head axis before the column-parallel O projection. Weights are stored once: GPU shards
keep their native SIMT layout, and only CPU shards are repacked offline at initialization
into the SME $16\times64$ tile layout, after which the temporary unpacked CPU copies are
freed; \sysname{} therefore holds exactly one copy of the model weights
(Figure~\ref{fig:uma_and_rearrange_weight}).}

However, realizing this zero-copy, in-place mutation fundamentally conflicts with the functional semantics of modern lazy computation graphs. In standard ML compilers, graph abstractions inherently conflate logical data dependence with physical memory materialization. Merging CPU and GPU sub-tensors strictly requires a \texttt{concat} operator, which is lowered into explicit memory allocations and copy kernels, thereby wasting the hardware benefits of UMA. Conversely, performing a naive in-place write into a shared buffer violates the DAG's tensor-level dependency tracking. Without explicit data-flow edges, the instruction scheduler cannot guarantee execution order, leading to severe data hazards (e.g., downstream operators reading stale data) unless penalized by conservative, eager synchronization.

To safely merge CPU and GPU results without triggering memory copies or breaking data dependencies, we modify the graph execution via two lightweight mechanisms:
(1) Logical Dependency Barrier: We replace the standard \texttt{concat} operator with a custom barrier node. Logically, it acts as a synchronization point in the DAG, ensuring downstream operators wait for both processors to finish. Physically, it allocates zero memory, simply aliasing the output pointer to the underlying shared buffer.
(2) Heterogeneous Fast Path: As the logical barrier handles the graph dependencies, we can execute this specific subgraph outside the framework's generic scheduling loop. This dedicated fast path issues concurrent tasks directly to the CPU and GPU without inserting conservative cross-device hardware fences, achieving a truly lock-free, zero-copy concurrent execution.

\revised{\paragraph{Coherence and memory-model assumptions.}
\sysname{} does not introduce a new coherence protocol. It relies on the coherence already provided by the UMA SoC: the CPU and GPU share a single physical address space and are kept coherent through the system-level cache, so a write issued by either device to a shared buffer is visible to both devices without explicit cache flushes or host–device copies. This single coherence domain is exactly the property that the concat-based path in stock MLX discards, since concat materializes a new contiguous array and copies both halves into it. Consistency, however, is weak rather than sequential: Metal follows a weak, relaxed memory model, where cross-device visibility is not guaranteed without explicit synchronization. Our Heterogeneous Fast Path therefore establishes only the producer–consumer happens-before relation the design actually needs. The MLP projections and the attention Q/K/V/O projections are all column-parallel, so the CPU and GPU write disjoint column ranges: no address is ever written by both devices, so there is no data race and no fine-grained per-line coherence is required. The Logical Dependency Barrier encodes the remaining ordering as a graph dependency, a zero-copy node that aliases the shared buffer, so MLX's lazy evaluation orders downstream operators after both producers finish, with no explicit hardware fences or eager synchronization. }

\subsubsection{SME Micro-kernel Optimization}
\label{sec:sme}

Maximizing the efficiency of the CPU requires saturating the SME units.
The system achieves this saturation through a tightly coupled hierarchy of micro-optimizations. 

At the single-core instruction level, we partition the matrix workloads into $16 \times 64$ micro-tiles. 
This specific dimension is purposefully chosen to simultaneously saturate the memory bandwidth using continuous multi-vector loads (\texttt{svld1\_x4}) and to keep all four 2D ZA register tiles maximally occupied. 
By aggressively unrolling loops along the K dimension, the kernel physically overlaps these continuous memory loads with the outer-product accumulation. 
This software pipelining effectively masks memory access latencies. 
To ensure these saturated compute pipelines are not abruptly starved by cache thrashing, we systematically tune the macro-tiling parameters. 
By strictly confining the active memory footprint within the 8MB effective L2 working set, the system avoids sharp bandwidth degradation and maintains continuous data ingestion.

However, scaling to full SoC utilization exposes load imbalance.
The asymmetric multiprocessing (AMP) topology of modern edge chips, which mixes high-throughput P-cores with slower E-cores, as illustrated in Figure~\ref{fig:uma_and_rearrange_weight}, inevitably induces straggler effects under traditional static task binding. 
To mitigate these straggler effects, we deploy a lock-free dynamic work-stealing mechanism. 
Built entirely upon relaxed-consistency atomic operations, this decentralized queue allows heterogeneous worker threads to autonomously steal compute micro-tiles based on their real-time physical execution rates. 
This achieves optimal fine-grained load balancing across the asymmetric hardware topology without ever incurring lock contention overheads.

\subsection{Agent-Aware Scheduling Layer }
\label{sec:scheduling_layer}

\begin{figure}[!t]
    \centering
    \includegraphics[width=0.98\linewidth]{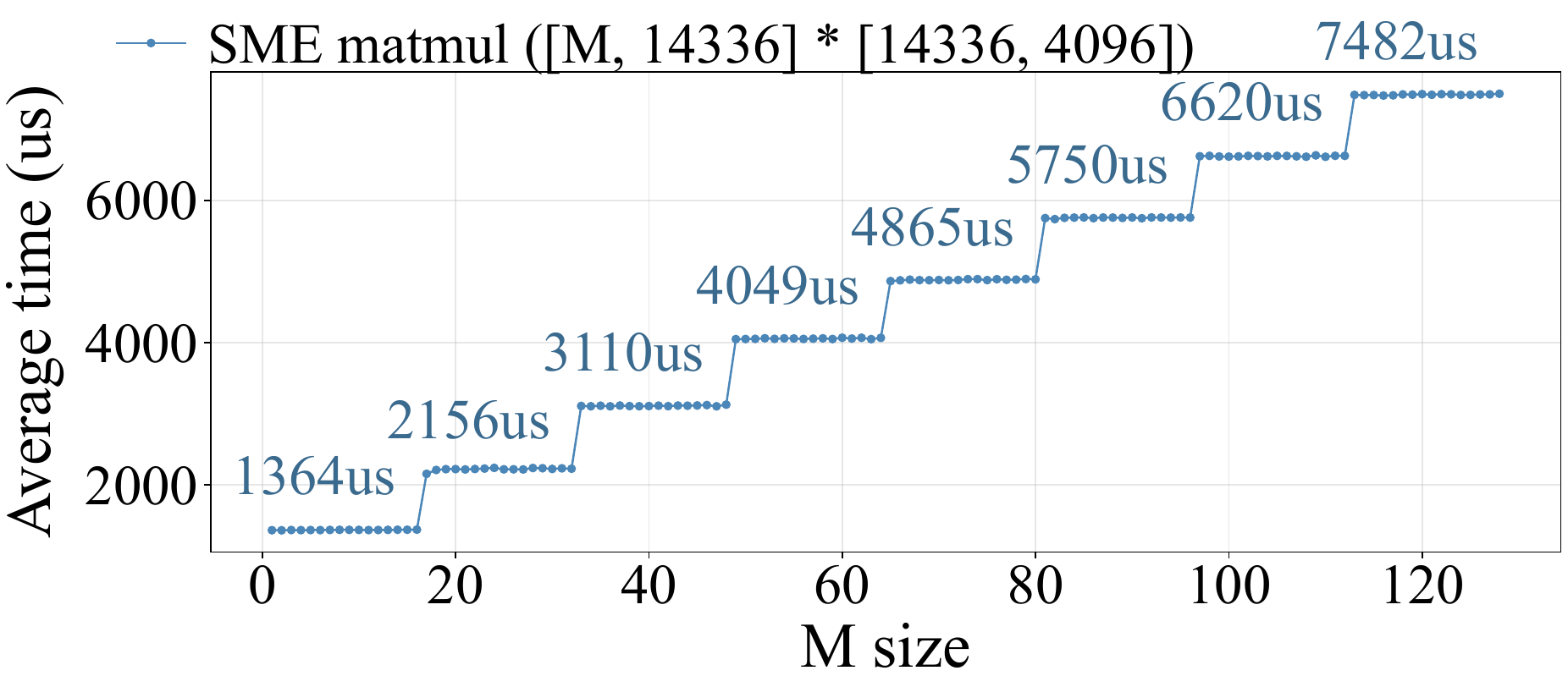}
    \caption{The stage performance of SME matmul.}
    \label{fig:sme_kernel}
\end{figure}

The scheduling layer acts as the bridge between multi-agent workflows and the physical hardware constraints. We first determine the optimal global drafting budget $M$ from hardware profiling results.
As shown in Figure~\ref{fig:sme_kernel}, the SME matrix multiplication kernel exhibits a strict stage performance behavior. 
Driven by $16 \times 64$ micro-tiles, the matmul performance scales in discrete jumps at batch size multiples of 16, with efficiency plateauing between these thresholds. 
Critically, beyond a draft length of 16, the performance degrades sharply, establishing an upper bound for efficient batch processing.
Furthermore, Figure~\ref{fig:spec_nonlinear} also validates this choice through end-to-end speculative decoding profiling.
For high-acceptance-rate tasks (Structured Output Task), the maximum speedup of 1.69$\times$ is achieved at a draft length of 16, after which performance begins to degrade due to rising rejection overhead. 
Therefore, to align with the hardware efficiency of both CPU and GPU units, we set the global drafting budget to $M=16$ token slots. Given this fixed hardware-aligned capacity, the scheduling layer employs two dynamic mechanisms to maximize the utility of these slots.

\subsubsection{Dynamic Drafting Budget Allocation}
\label{sec:dynamic_spec}

To reliably saturate these 16 hardware slots without forcing a single request to over-expand its speculative tree (which suffers from diminishing returns in acceptance yield), the system pools multiple concurrent agents into a unified physical batch. However, uniformly dividing this fixed 16-token budget fails to account for the extreme variance in the difficulty of drafting across streams. Static allocation strategies inherently bottleneck parallel efficiency by over-provisioning execution slots to difficult tasks while under-resourcing highly productive ones. To dynamically allocate draft length based on these runtime characteristics, we propose an \textit{agent-aware scheduler}. By introducing the Historical Accepted Length (HAL) as an objective proxy for the underlying drafting difficulty, this scheduler seamlessly captures the real-time generation state of each concurrent stream.

The HAL tracks the absolute number of accepted tokens rather than the acceptance rate. Relying on percentages obscures an agent's true throughput contribution, as a shallow draft tree might exhibit a perfect acceptance rate while yielding minimal tokens. \revised{HAL is therefore a lightweight heuristic rather than a globally optimal metric: it steers the fixed $L_{total} = 16$ slots toward agents that yield more accepted tokens per verification step.} After the $t$-th verification step, the scheduler applies an exponential moving average (EMA) to update the HAL for each active agent $i$:
$$A_i(t) = \gamma \cdot k_i(t) + (1 - \gamma) \cdot A_i(t-1)$$
where $k_i(t)$ represents the absolute count of accepted tokens in the current step. The decay factor $\gamma$ dictates the responsiveness to sudden shifts in drafting difficulty within the generative stream. 
In practice, we perform an exhaustive grid search over candidate hyperparameters using profiling data from heterogeneous multi-agent workloads. 
The decay factor $\gamma = 0.3$ and initial prior of $A_i(0) = 4.0$ are selected as the optimal combination that maximizes end-to-end system throughput, effectively balancing responsiveness to workload variations and noise robustness.

For the subsequent decoding step $t+1$, the scheduler must partition the global drafting budget $L_{total}=16$ across $N$ concurrent agents. 
To structurally guarantee starvation-free forward progress, the system provisions a non-preemptible baseline capacity, $L_{min}$, for each agent. The residual budget is then dynamically allocated using a temperature-scaled softmax distribution over the agents' HAL:
$$L_i(t+1) = L_{min} + \left\lfloor (L_{total} - N \cdot L_{min}) \cdot \frac{\exp(A_i(t) / \tau)}{\sum_{j=1}^N \exp(A_j(t) / \tau)} \right\rfloor$$
This agent-aware control loop autonomously skews the draft length allocation toward high-yield agents. Consequently, it physically truncates the speculative draft trees of difficult-to-draft workloads, strictly bounding non-productive compute exploration and memory overheads based purely on empirical acceptance signals. \revised{Allocation is bounded on both sides by this formula. Because every agent reserves the non-preemptible $L_{min}$ first, the softmax only distributes the residual $L_{total} - N * L_{min}$ , so even the highest-HAL agent cannot capture the full $L_{total}=16$ budget. Because HAL is recomputed through the EMA at every verification step, a transient burst cannot permanently lock out other agents.}

\subsubsection{Cooperative Suspend-and-Yield Mechanism}
\label{sec:suspend_yield}

External tool invocations inevitably stall individual execution streams. In a strictly hardware-aligned concurrent batch, a stalled agent occupying active token slots will directly degrade global throughput. To explicitly neutralize these latencies, the \textit{agent-aware scheduling layer} implements an OS-inspired cooperative suspend-and-yield mechanism tied directly to the speculative budgeting system.

At the synchronization boundary of each verification step, the scheduler inspects agent execution states. Upon detecting a tool-induced stall, it triggers a cooperative yield, migrating the stalled agent to a suspended queue and instantly relinquishing its allocated token slots back to the global drafting budget. Relinquishing these slots allows the dynamic allocator (Section~\ref{sec:dynamic_spec}) to instantly redistribute the yielded budget across the remaining active agents. By absorbing the vacated budget, active streams dynamically expand their draft trees, broadening their speculative search space without breaking the optimal hardware-aligned 16-token threshold. 

Crucially, by exploiting the physically shared addressing of the UMA architecture, the system performs an in-place metadata freeze on the suspended agent's KV cache. Upon resolution of the external tool execution, the scheduler promotes the ready agent back to the active queue to request fresh token slots. As the underlying memory pool guarantees in-place KV freezing and hardware-aware offset-remapping compaction, the awakened agent directly resumes speculative execution. This tightly integrated memory management explicitly eliminates the need for expensive context re-prefill phases, achieving strictly zero-overhead state restoration.

\section{Evaluation}
\sysname{} builds upon Apple's native MLX runtime, fundamentally restructuring its underlying C++ graph executor to enable zero-copy UMA memory management. Furthermore, it introduces lightweight cross-layer APIs to synchronize multi-agent execution states such as variance in drafting difficulty and tool-induced stalls with the underlying scheduler. We comprehensively evaluate \sysname{} using trace-driven multi-agent workflows to demonstrate its end-to-end makespan, concurrent throughput, and resilience to tool-induced execution stalls.

\subsection{Experimental Setup}

\textbf{Hardware Testbed.}
Primary experiments are conducted on an Apple M4 SoC equipped with 32\,GB of unified memory and a 120\,GB/s UMA bus. The SoC integrates a 10-core CPU (4 performance + 6 efficiency) with the ARMv9 SME2 extension, alongside a 10-core GPU. \revised{To verify \sysname{}'s generalization, the key ablations are additionally evaluated on an Apple M4 Pro SoC with 64\,GB of unified memory and a 273\,GB/s UMA bus, whose 14-core CPU (10 performance + 4 efficiency) and 20-core GPU shift the compute-to-bandwidth balance relative to the base M4. Since both SoCs share the SME2 extension, the M4 Pro isolates the effect of the bandwidth balance rather than the micro-kernel backend. Together they exemplify the stringent memory-bandwidth and computational constraints of modern edge AI environments.}

\textbf{Models.}
We evaluate our system using two representative large language models: DeepSeek-R1-Distill-Llama-8B~\cite{deepseek2025r1}, a reasoning-focused model that exhibits severe drafting difficulty (yielding low acceptance rates) during its extensive chain-of-thought (CoT) generation, and Llama-3.1-8B-Instruct~\cite{grattafiori2024llama} for standard instruction-following workloads. For configurations involving speculative decoding, we uniformly adopt EAGLE-3~\cite{li2025eagle3}, a state-of-the-art draft generation and verification framework. All models and tensor operations are executed in FP16 precision.

\textbf{Datasets and Workloads.}
Lacking deterministic multi-agent benchmarks for rigorous micro-architectural evaluation, we construct a workload-representative trace synthesizer. We first sample real-world prompts to build a heterogeneous workload pool across three distinct agent domains, whose profiling characteristics are summarized in Table~\ref{tab:dataset_profiling}:

\begin{itemize}[leftmargin=*]
\item \textbf{Reasoning Task (LongBench~\cite{bai2025longbench}):} Features hard-to-draft logical reasoning, such as orchestrator making decisions and reading comprehension summaries.
\item \textbf{Structured Output Task (MBPP~\cite{austin2021program}):} Comprises programming tasks that generate structured, highly predictable code sequences.
\item \textbf{Tool Execution Task (ToolBench~\cite{guo2024stabletoolbench}):} Provides tool-use and API invocation prompts, representing the interactive behaviors of autonomous agents.
\end{itemize}

To construct the synthetic trace, our workload generator applies three steps. First, we blend the sampled prompts in a 1:2:2 ratio (Reasoning : Code : Tool). \revised{This reflects the structure of real agent systems, where a reasoning orchestrator is outnumbered by code- and tool-executing workers. It also avoids the degenerate reasoning-heavy regime: reasoning agents accept few tokens per verification step and therefore take many steps, so a 1:1:1 mix ends in a long tail in which a single reasoning agent drains the queue alone, the speculative batch is under-filled, and no scheduling headroom remains for HAL to redistribute.} Second, we inject log-uniform stochastic pauses into tool-use requests to simulate API latencies. Finally, to evaluate realistic scheduling and makespan, requests are dispatched via a Poisson process with intra-agent delays; conversely, peak throughput measurements bypass these arrival delays to maintain continuous system saturation.

\begin{table}[!t]
\caption{Characteristics of the sampled tasks used to construct the steady-state workload pool. We profile the average prompt length and speculative acceptance length for each dataset under a fixed draft length of $L=16$.}
\label{tab:dataset_profiling}
\centering
\small
\setlength{\tabcolsep}{3pt}
\begin{tabular}{l|c|cc}
\toprule
\multirow{2}{*}{\textbf{Dataset}} & \multirow{2}{*}{\begin{tabular}[c]{@{}c@{}}\textbf{Avg Prompt}\\ \textbf{Length}\end{tabular}} & \multicolumn{2}{c}{\textbf{Avg Accept Length ($L=16$)}} \\ \cmidrule(l){3-4}
 & & \textbf{LLaMA-3.1-8B} & \textbf{DeepSeek-R1-8B} \\ \midrule
LongBench & 3950 & 1.19 & 1.24 \\
MBPP      & 36   & 2.67 & 6.33 \\
ToolBench & 114  & 2.81 & 4.98 \\ \bottomrule
\end{tabular}
\end{table}

\textbf{Baselines.}
To systematically isolate our performance contributions, we evaluate \sysname{} against three baseline configurations, all executed strictly on the GPU to represent standard state-of-the-practice deployments:
\begin{itemize}[leftmargin=*]
\item \textbf{Batch-AR:} Vanilla non-speculative baselines for naive continuous batching.
\item \textbf{Seq-SD (EAGLE-3):} State-of-the-art single-stream speculative decoding. It establishes the isolated single-agent performance bound.
\item \textbf{Batch-SD (Batched EAGLE-3):} The strongest concurrent baseline, utilizing GPU-bound continuous batching with a fixed per-request draft budget ($L_i=16$). It highlights two critical flaws of standard scheduling: computational waste from static draft lengths, and severe compute stalling during agent tool execution.
\end{itemize}

\textbf{Metrics.} We evaluate overall performance across two primary dimensions: user-centric latency and system-centric throughput. To capture the actual user-perceived turnaround time, we first measure the end-to-end makespan, defined as the total wall-clock time required to complete the hybrid workload. Complementarily, the system throughput is measured by global throughput, the aggregate token generation rate across all concurrent streams.

\subsection{End-to-End Evaluation}
\label{sec:eval_e2e}

\begin{figure*}[!t]
    \centering
    \includegraphics[width=0.98\linewidth]{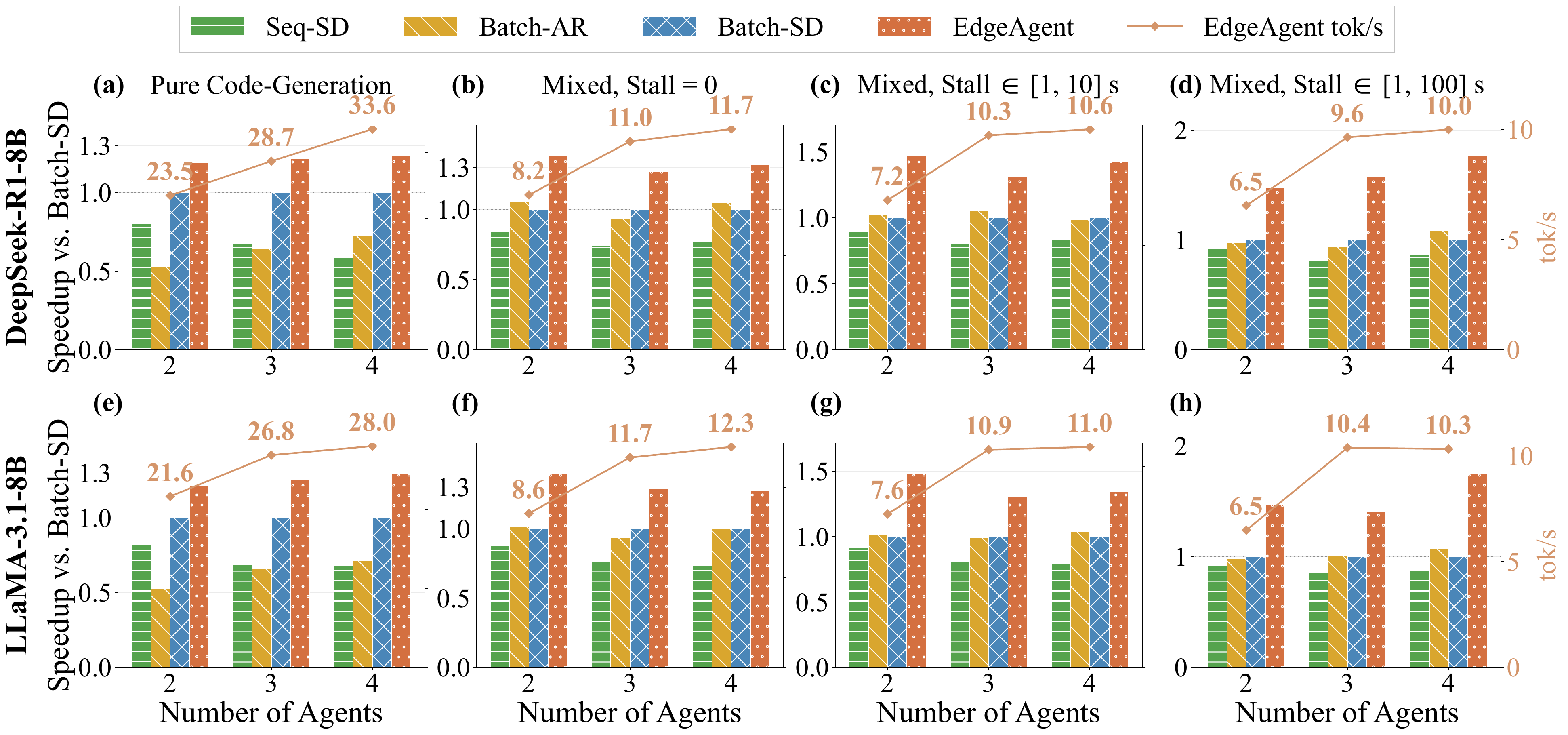}
    \caption{\reviseded{End-to-end evaluation of \sysname{} against baselines under varying concurrency levels. The left axis (bars) shows the makespan speedup over Batch-SD, while the right axis (line) details the absolute global throughput of \sysname{}. Top (a-d) and bottom (e-h) rows evaluate DeepSeek-R1-8B and LLaMA-3.1-8B. Columns from left to right transition from pure code-generation to mixed reasoning/code/tool-generation workloads with increasing tool-induced stall durations.}}
    \label{fig:e2e_evaluation}
\end{figure*}

\begin{figure}[!t]
    \centering
    \includegraphics[width=\linewidth]{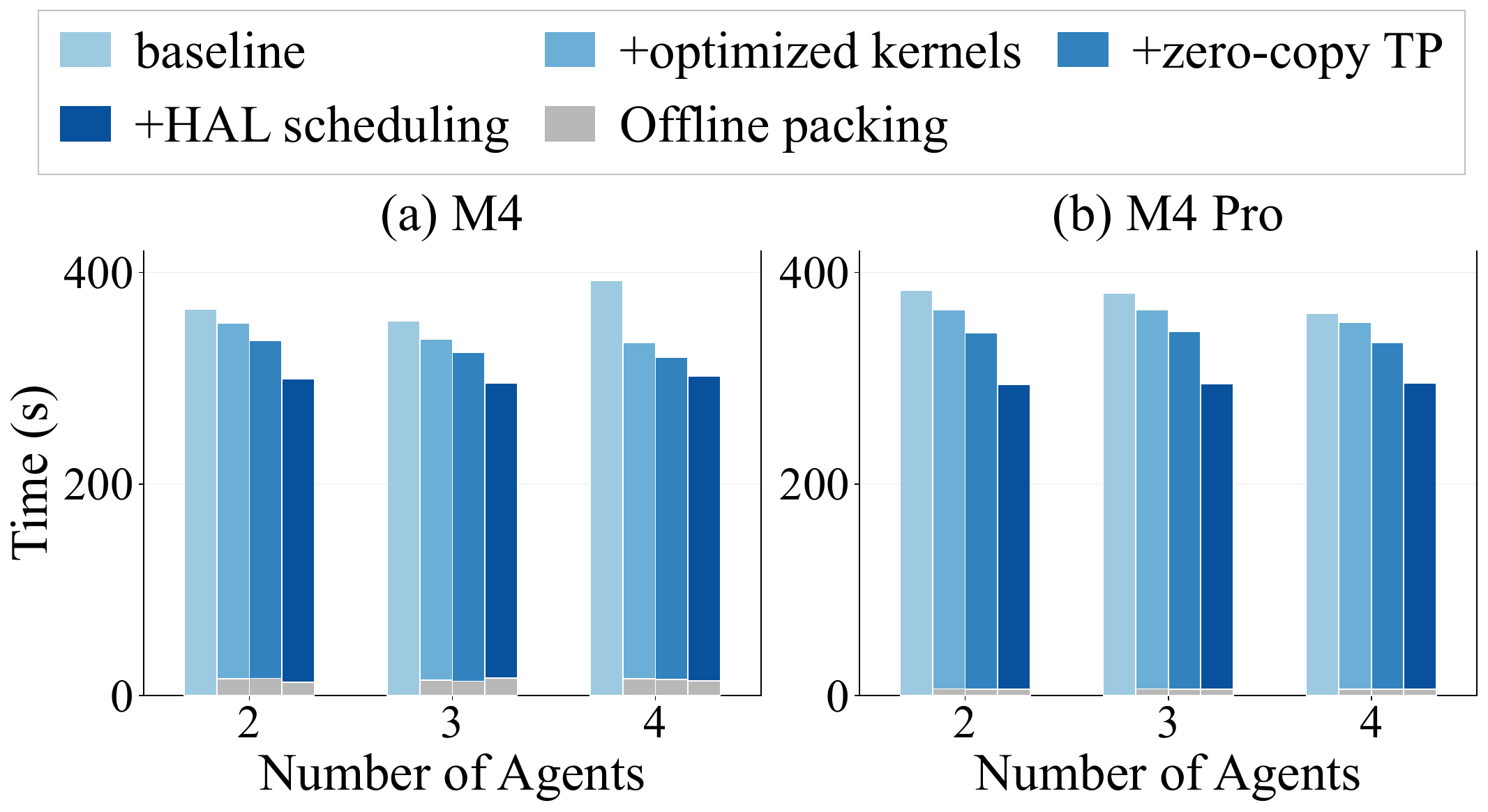}
    \caption{\revised{Component ablation of the end-to-end makespan on both the M4 and the M4 Pro.}}
    \label{fig:detailed_ablation}
\end{figure}

\begin{figure}[!t]
    \centering
    \includegraphics[width=\linewidth]{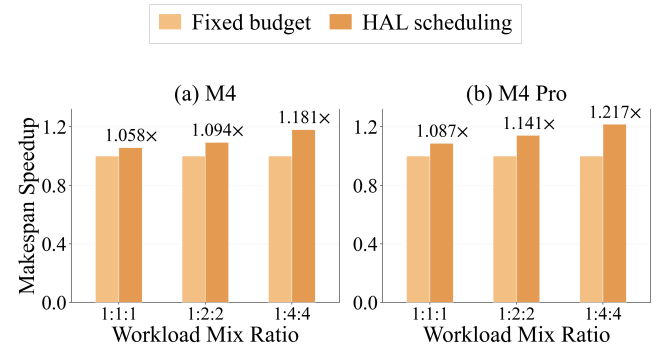}
    \caption{\revised{Sensitivity of HAL scheduling to the reasoning/code/tool workload mix.}}
    \label{fig:diferent ratio workload}
\end{figure}

To evaluate the overall makespan and global throughput of \sysname{}, we construct a spectrum of concurrent workloads. 
As illustrated in Figure~\ref{fig:e2e_evaluation}, we progressively inject task heterogeneity and external tool-induced stalls to expose the scheduling bottlenecks of existing frameworks. This evaluation isolates the end-to-end efficiency of the system, while the specific micro-architectural mechanisms driving these results are analyzed in the subsequent subsections  (\S\ref{sec:eval_tp}--\S\ref{sec:eval_io}).

\paragraph{Baseline Inference Performance.} 
We first establish the baseline performance using a continuous, pure code-generation workload without tool-induced stalls (Figures~\ref{fig:e2e_evaluation}(a) and \ref{fig:e2e_evaluation}(e)). At $N=4$, where $N$ denotes the number of actively decoding agents in the system, \sysname{} achieves a global throughput of 33.6 tok/s (DeepSeek) and 28.0 tok/s (LLaMA). 
\revised{These metrics represent $1.29\times$ speedups over the optimized Batch-SD baseline. This gain reflects our UMA-aware execution layer, which combines the optimized SME kernels with zero-copy TP, prior to enabling HAL scheduling and stall-aware context switching. }

\revised{\paragraph{Component ablation.}
To isolate each contribution, we enable EdgeAgent's optimizations incrementally on top of the Batch-SD baseline: first the optimized SME kernels, then zero-copy TP, and finally HAL scheduling, measuring the end-to-end makespan for 2, 3, and 4 concurrent agents on both the M4 and the M4 Pro (Figure~\ref{fig:detailed_ablation}). Each step reduces the makespan. Together, the optimized SME kernels and
zero-copy TP constitute our UMA-aware execution layer, which achieves up
to a $1.29\times$ speedup over Batch-SD. Enabling HAL scheduling provides
an additional $1.05$--$1.17\times$ improvement over the corresponding
UMA-aware configuration. Offline packing is included in every bar and is small: 13–16 s on the M4 and 6.2 s on the M4 Pro, under $5\%$ of the makespan. }

\paragraph{Adaptability to Mixed Workload.} 
To evaluate scheduling efficiency under varying sequence predictability, we introduce a mixed workload combining reasoning and coding tasks without tool-induced stalls (Figures~\ref{fig:e2e_evaluation}(b) and \ref{fig:e2e_evaluation}(f)). While the overall throughput naturally decreases across all systems due to the hard-to-draft reasoning steps, a static draft-length allocation ($L=16$) wastes compute resources generating and
verifying unaccepted tokens during low-acceptance reasoning phases. By utilizing dynamic budgeting (\S\ref{sec:eval_sched}), \revised{\sysname{} mitigates this overhead and achieves a 1.33$\times$ speedup over the Batch-SD baseline}, demonstrating the efficiency of predictability-aware resource allocation.

\revised{\paragraph{Sensitivity to Workload Composition.}
To test whether the 1:2:2 default mix flatters our results, we sweep the reasoning/code/tool ratio from 1:1:1 to 1:4:4 while holding the arrival process and concurrency settings fixed. HAL scheduling improves on the fixed-budget baseline across the entire sweep, and its gain grows monotonically as the mix shifts toward structured code and tool generation
(Figure~\ref{fig:diferent ratio workload}). This is the expected direction: structured-output
agents are highly predictable and a static $L=16$ budget caps the tokens they could otherwise accept, whereas extra budget spent on reasoning-heavy agents is largely wasted on rejected drafts. The 1:2:2 default therefore sits at an interior and conservative operating point of this curve, not a favorable peak.}

\paragraph{Adaptability to Tool Stalls.} 
Real-world agent workflows frequently stall during external tool invocations. To simulate this, we inject log-uniform tool-execution stall durations of $[1, 10]$ seconds and $[1, 100]$ seconds into the mixed workload (Figures~\ref{fig:e2e_evaluation}(c)-(d) and \ref{fig:e2e_evaluation}(g)-(h)). Under the extreme $[1, 100]$s stall configuration at $N=4$ (Figure~\ref{fig:e2e_evaluation}d), head-of-line blocking in Batch-SD severely delays execution, inflating its makespan to 213.1 seconds (DeepSeek). In contrast, \sysname{} leverages its asynchronous suspend-and-yield mechanism (\S\ref{sec:eval_io}) to evict stalled agents and reclaim verification slots. This effectively hides the stall, compressing the total makespan to just 120.6 seconds and delivering a 1.77$\times$ speedup over the optimized Batch-SD baseline.

\paragraph{Overall Performance Scaling.} 
Across the evaluated configurations, our UMA-aware execution layer,
which combines the optimized SME kernels with zero-copy TP, achieves up
to a $1.29\times$ speedup over Batch-SD. The primary systemic gains emerge
in stall-intensive environments, where the complete \sysname{} system
reaches a $1.77\times$ speedup by combining UMA-aware execution, HAL
scheduling, and stall-aware context switching. Furthermore, scaling concurrency from $N=2$ to $N=4$ consistently widens this margin, indicating that the system effectively utilizes edge UMA resources under highly irregular execution patterns.

\subsection{Evaluation on Tensor Parallelism}
\label{sec:eval_tp}

\begin{figure}[!t]
    \centering
    \includegraphics[width=\linewidth]{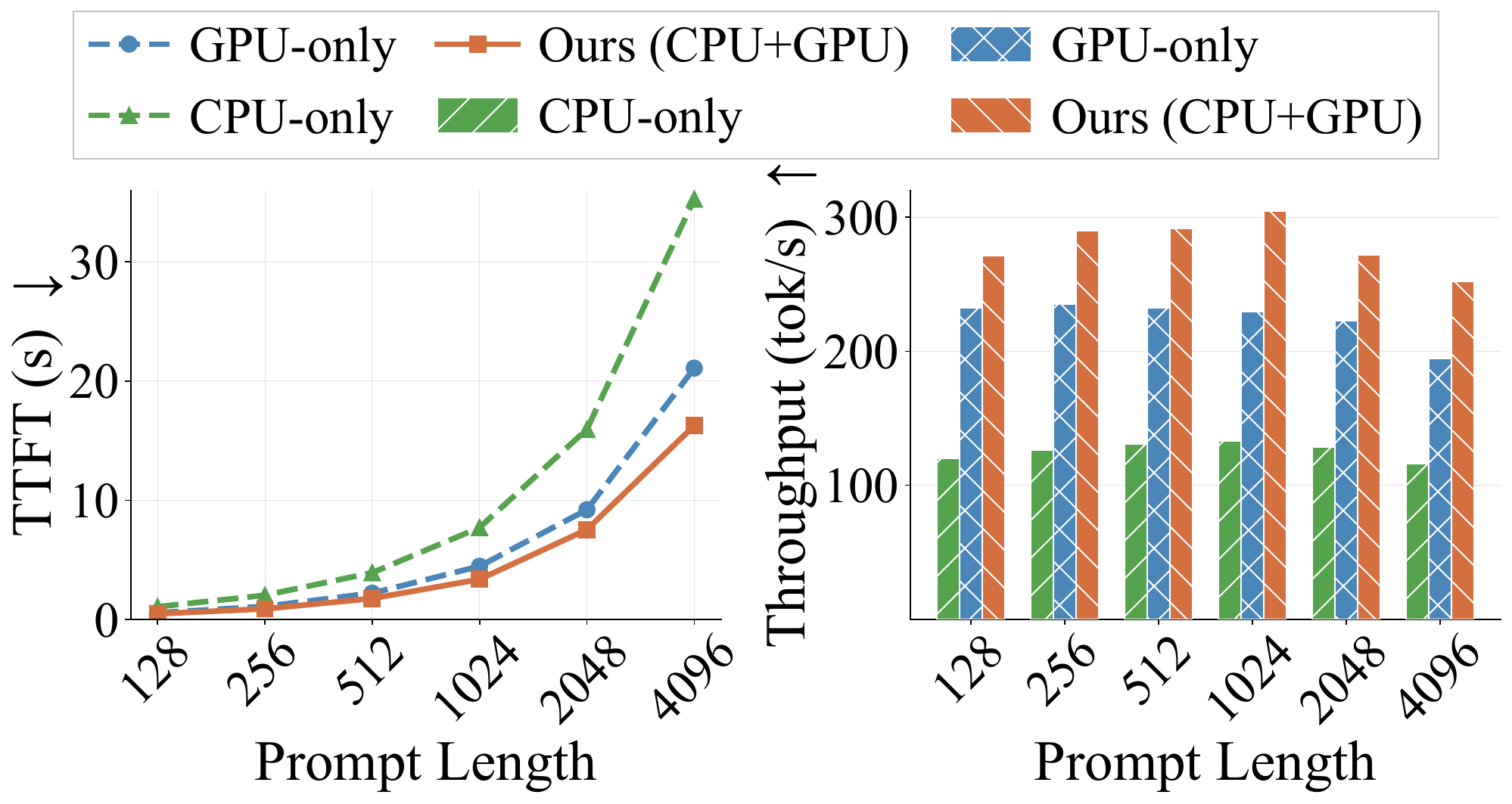}
    \caption{Prefill performance scaling across varying prompt lengths. The figure illustrates the time to first token (TTFT) and absolute processing throughput for standalone processors versus our heterogeneous co-execution.}
    \label{fig:prefill_performance}
\end{figure}

\begin{figure}[!t]
    \centering
    \includegraphics[width=\linewidth]{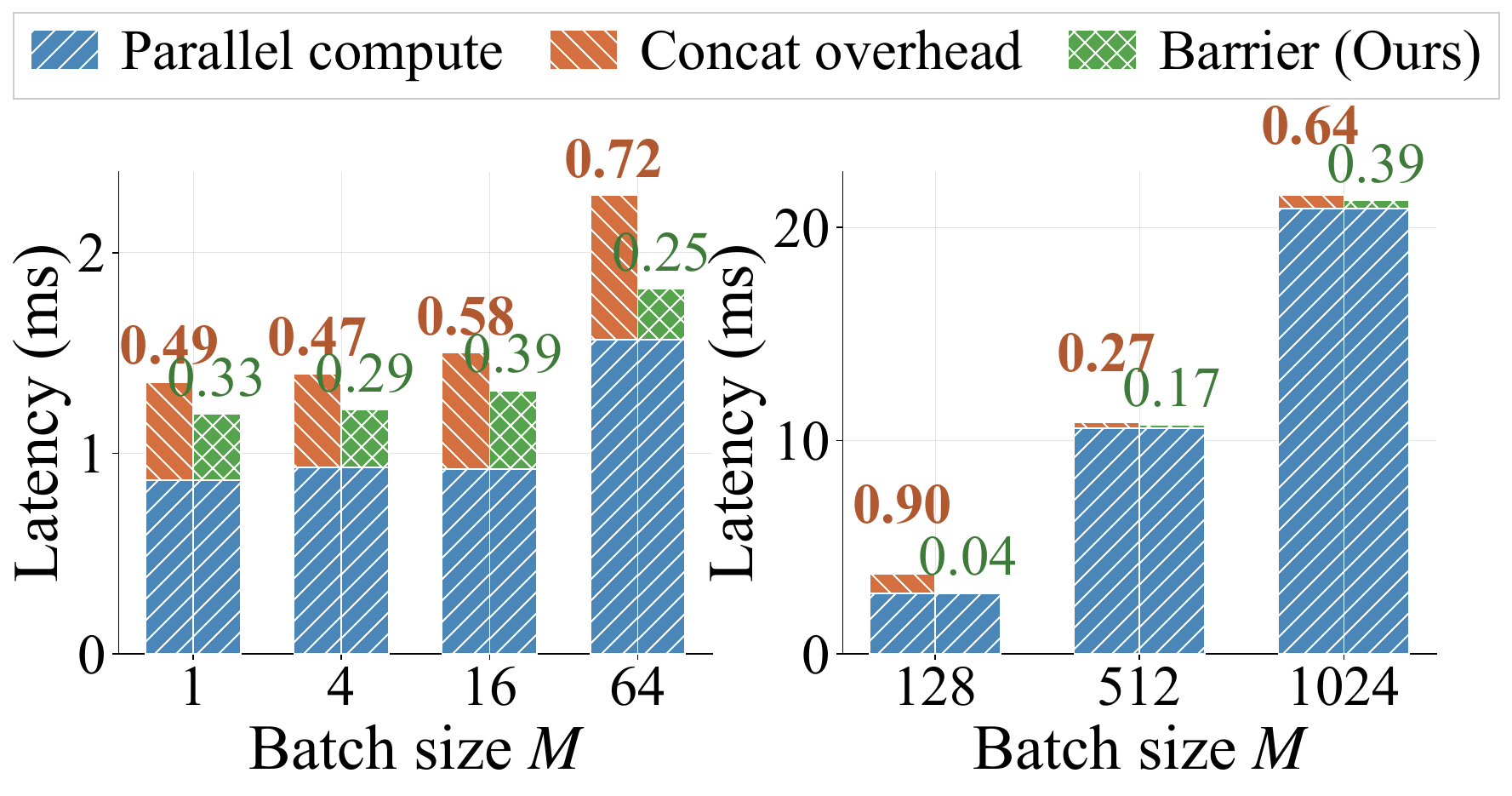}
    \caption{Latency decomposition of UMA co-execution across varying batch sizes $M$ on a $4096 \times 14336$ projection. \sysname{} replaces the Concat memory copy shown in orange with a lightweight zero-copy barrier shown in green.}
    \label{fig:latency_decomp}
\end{figure}

\begin{figure}[!t]
    \centering    
    \includegraphics[width=\linewidth]{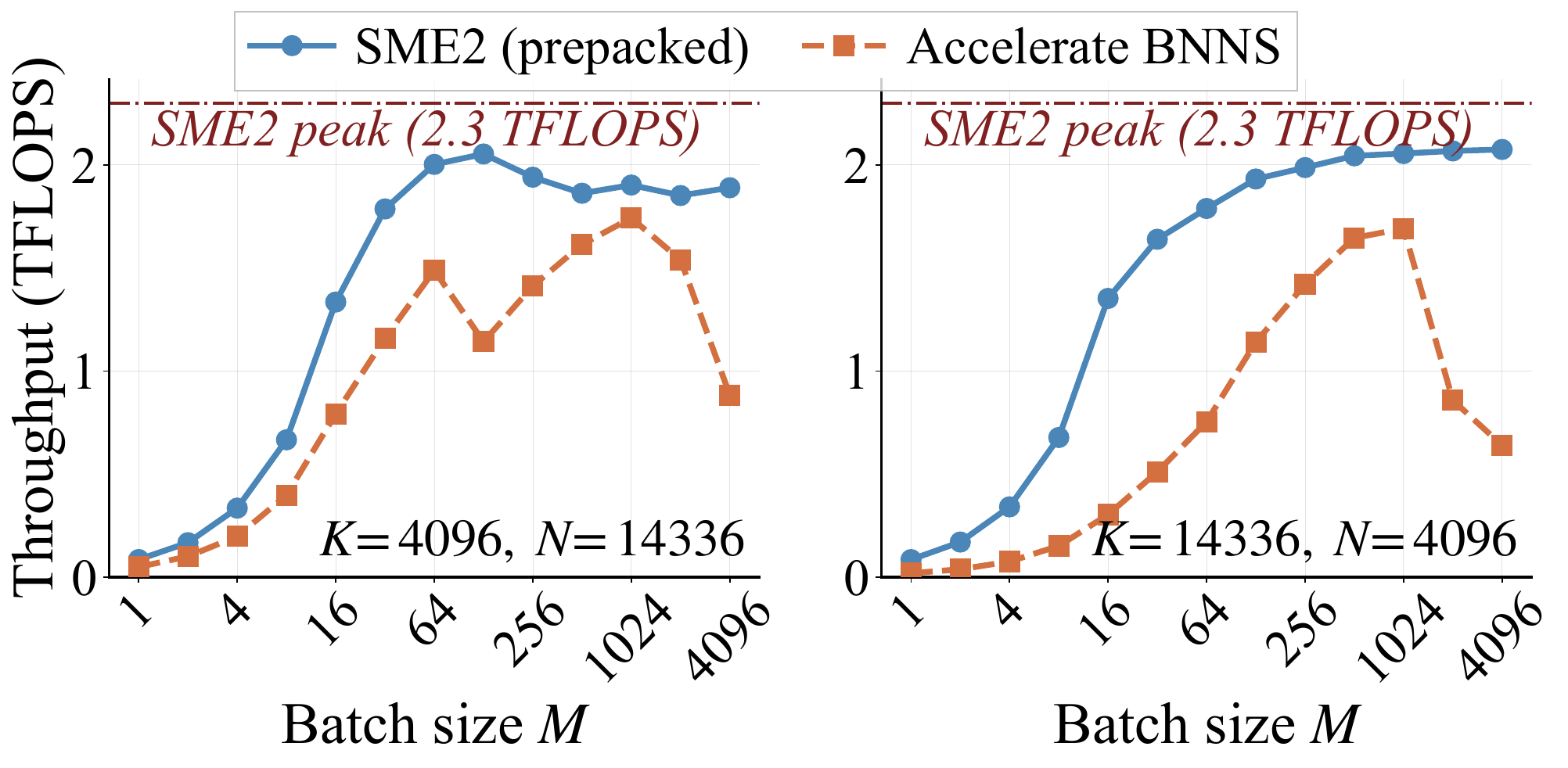}
    \caption{Performance comparison measured in TFLOPS between Apple's BNNS library and our custom SME kernel across varying batch sizes $M$. The SME kernel consistently approaches the hardware's theoretical peak of $\sim$2.3 TFLOPS.}
    \label{fig:eval_sme_kernel}
\end{figure}

The foundational execution engine of \sysname{} is built upon our architecture-aware heterogeneous tensor parallelism. This section systematically dissects how the framework maximizes the absolute compute capacity of the UMA architecture while mitigating the synchronization overhead typically associated with cross-device execution. We begin by quantifying the macroscopic throughput and latency improvements achieved during the compute-bound prefill phase. Following this, a rigorous latency decomposition demonstrates the elimination of tensor aggregation overhead via our zero-copy synchronization computation graph. The evaluation concludes by isolating the micro-architectural efficiency of our custom SME kernel, confirming its capacity to sustain hardware-saturating utilization across highly skewed inference workloads.

\subsubsection{Accelerating Compute-Bound Prefill}

To evaluate the prefill performance of our heterogeneous tensor parallelism, we adopt two metrics. The primary latency metric is time to first token (TTFT), which measures the end-to-end duration required to process the input prompt and generate the initial token. To quantify overall system capacity, we measure prefill throughput, defined as the number of prompt tokens processed per second. 
They demonstrate the efficiency of our CPU-GPU co-execution under large contextual workloads, or rather the compute-bound workload.

The results presented in Figure~\ref{fig:prefill_performance} confirm the compute-bound nature of the prefill phase, where standalone GPU and CPU executions remain stable at 232 and 132 tokens per second. By leveraging heterogeneous tensor parallelism, our system effectively surpasses these isolated hardware ceilings. At a prompt length of 1024, the co-execution engine achieves a peak throughput of 304.5 tokens per second, yielding a 1.31$\times$ speedup over the GPU-only baseline. This compute expansion directly translates to latency improvements for extended contexts. For a 4096 token prompt, our system reduces the TTFT from 21.0 seconds on the standalone GPU to 16.2 seconds. These measurements demonstrate that our architecture successfully aggregates physical compute units without triggering shared memory bus contention during dense matrix operations.

\subsubsection{Zero-Copy Synchronization}

Standard UMA co-execution relies on explicit tensor concatenation \texttt{Concat}, triggering expensive $\mathcal{O}(M \cdot N)$ memory copies that bottleneck the shared bus. \sysname{} eliminates this via a zero-copy computation graph, pre-allocating a single memory block for in-place, simultaneous writes and reducing aggregation to a lightweight barrier. 

Figure~\ref{fig:latency_decomp} decomposes the end-to-end latency.
In the memory-bound decode phase ($M \in \{1, 4, 16\}$), baseline memory copies (\texttt{Concat}) inflate the sub-millisecond pure compute time by >50\%. \sysname{}'s zero-copy architecture eliminates this bottleneck, yielding a consistent $\sim$1.14$\times$ speedup. At micro-batch $M=128$, the overhead remains severe (+0.90~ms on top of $\sim$2.8~ms compute), which \sysname{} compresses to just +0.04~ms, delivering a 1.31$\times$ speedup. Conversely, during massive prefill ($M=1024$), the dense $\mathcal{O}(M \cdot K \cdot N)$ compute surges past 20~ms, naturally amortizing the $\mathcal{O}(M \cdot N)$ synchronization overhead to $<3\%$. This confirms that \sysname{} introduces negligible synchronization delays across all batch scales.

\subsubsection{Custom SME Kernel efficiency}

To maximize CPU utilization during co-execution, relying on general-purpose libraries like Apple's BNNS is suboptimal for the highly skewed tensor shapes of LLM inference. We address this by implementing a custom GEMM kernel strictly optimized for ARM's SME. Unlike stateless BLAS libraries that redundantly repack static weights upon invocation, our kernel pre-packs model weights offline during initialization. This eliminates data-layout transformation overheads from the runtime critical path.

We evaluate our kernel against BNNS on standard MLP projections, specifically the $4096 \times 14336$ up-projection and the $14336 \times 4096$ down-projection. Measuring pure online inference throughput exclusive of offline packing time, the empirical results presented in Figure~\ref{fig:eval_sme_kernel} demonstrate major architectural advantages across both decoding and prefill regimes. During autoregressive decoding for $M \le 16$, BNNS suffers from significant dispatch overheads. For a batch size of $M=1$ on the down-projection layer, BNNS yields only 19.0 GFLOPS. By circumventing these framework overheads, our custom kernel delivers 85.0 GFLOPS, achieving a 4.47$\times$ speedup. Conversely, under massive prefill workloads for $M \ge 2048$, BNNS exhibits severe throughput degradation, dropping to 639.3 GFLOPS on the down-projection at $M=4096$. In contrast, our SME kernel maintains asymptotic stability with a peak throughput of 2.07 TFLOPS. This represents over 90\% of the 2.3 TFLOPS theoretical hardware limit, outperforming BNNS by 3.24$\times$ and ensuring the CPU effectively matches the GPU's computational pace.

\subsection{Dynamic Speculative Scheduling}
\label{sec:eval_sched}

\begin{figure}[!t]
    \centering
    \includegraphics[width=0.98\linewidth]{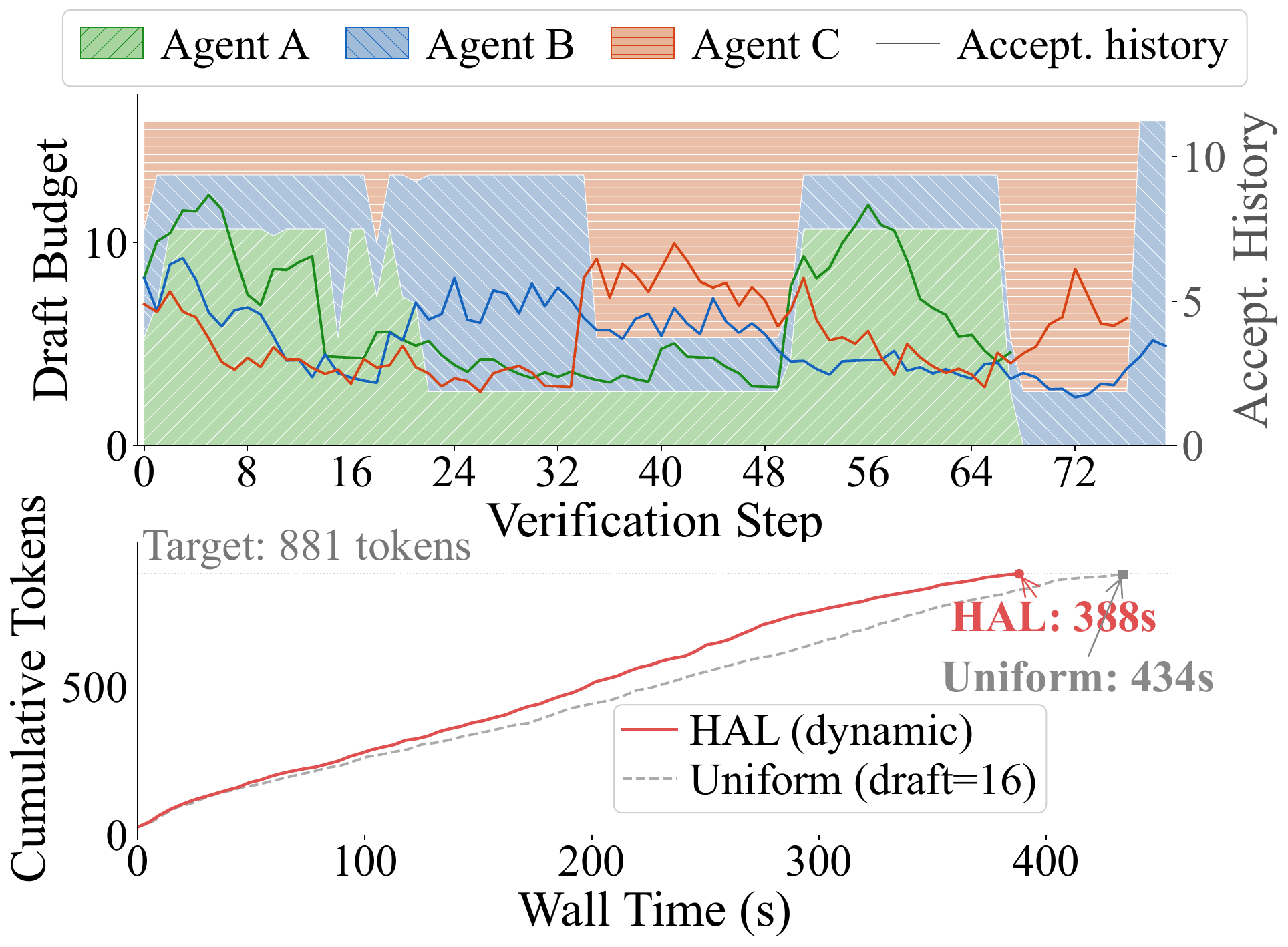}
    \caption{Evaluation of the HAL dynamic scheduler for concurrent multi-agent workloads. The upper panel shows per-agent HAL histories guiding drafting budget allocation. The lower panel demonstrates that the dynamic drafting budget allocation enables a higher throughput of a $1.12\times$ speedup over the static scheduling.}
    \label{fig:hal_dynamic}
\end{figure}

\begin{figure*}[!t]
    \centering
    \includegraphics[width=0.98\linewidth]{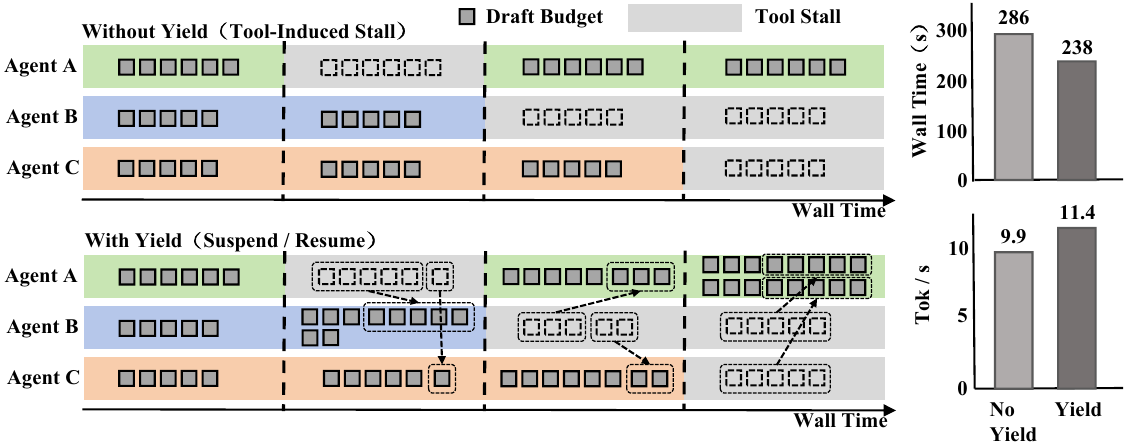}
    \caption{\revised{Execution trace of a three-agent workload. 
    The synchronous baseline (upper) indicates the severe hardware idling induced by tool execution, while our asynchronous yielding (lower) allows pending agents to steal idle budgets via microsecond-level context switches, thereby reducing wall-clock time and elevating system throughput (right panels).}}
    \label{fig:io_switch}
\end{figure*}

Traditional speculative decoding frameworks inherently underutilize resources by rigidly partitioning compute among agents, regardless of their actual generation predictability. 
To expose this inefficiency, we evaluate the dynamic scheduler with three concurrent agents, where each agent’s task is randomly sampled from the mixed dataset, thereby exhibiting differences in drafting difficulty. 
As shown in the execution trace of Figure~\ref{fig:hal_dynamic}, the system continuously monitors each agent’s Historical Accepted Length (HAL) using an exponential moving average. When an agent exhibits a sustained high acceptance trend, the scheduler reduces the draft budgets of low-acceptance agents and reallocates reclaimed hardware budgets to accelerate the most predictable ones. This highly responsive mechanism ensures that the autoregressive drafting phase perfectly mirrors the underlying workload dynamics.

Under the static and uniform budget allocation, hardware budgets are frequently wasted generating deep draft sequences for unpredictable, hard-to-draft workloads, which are ultimately rejected during the parallel verification phase. By preemptively suppressing these unproductive drafts and allocate draft budget to high-yield targets, our scheduler maximizes the effective accepted tokens per verification step. The lower panel of Figure~\ref{fig:hal_dynamic} illustrates this efficiency gain by plotting cumulative generated tokens against wall-clock time. By eliminating speculative computational waste, the dynamic policy achieves a strictly steeper generation trajectory compared to the baseline. For an identical workload output of $881$ target tokens, this continuous reallocation reduces the total execution time from $434$ seconds to $388$ seconds. This translates to a $1.12\times$ end-to-end speedup, demonstrating that HAL-guided dynamic budget allocation is fundamentally critical for optimizing concurrent multi-agent inference.

\subsection{Stall-Aware Context Switching}
\label{sec:eval_io}

To evaluate system resilience against tool-induced stalls, we benchmark our asynchronous yielding mechanism using three concurrent agents. Under standard synchronous execution, tool invocations critically block active execution streams. As depicted in the upper timeline of Figure~\ref{fig:io_switch}, waiting for external processes (such as script execution and API responses) forces the hardware into extended idle states, clearly illustrated by Agent B suffering a prolonged 53-second stall. These synchronous execution stalls entirely serialize external latency and tensor computation, severely degrading overall system efficiency.

Our stall-aware runtime eliminates this computational waste. Upon initiating an external tool request, the active agent suspends its execution context and yields the hardware. As shown via the interleaved scheduling paths in the lower timeline of Figure~\ref{fig:io_switch}, when Agent B blocks, the global scheduler instantly promotes Agent A or Agent C. Allowing companion agents to steal these idle budgets achieves deep overlap between tensor computation and tool execution. As quantified in the right panels of Figure~\ref{fig:io_switch}, this dense execution packing reduces the end-to-end wall-clock time from 286 to 238 seconds and elevates global throughput from 9.9 to 11.4 tokens per second, ensuring continuous hardware saturation despite highly irregular tool invocations.
\section{Related Work}

\noindent\textbf{LLM Serving for Agentic Workloads.}
The shift to ReAct-style agents~\cite{yao2023react} shatters the continuous batching assumptions of traditional runtimes like vLLM~\cite{vllm2024github}. To support complex control flows, SGLang~\cite{sglang2024github} and AgentInfer~\cite{lin2025towards} co-design runtime memory with application semantics, while Continuum~\cite{li2025continuum} and INFERCEPT~\cite{abhyankar2024infercept} mitigate tool-induced stalls via cache retention and KV swapping. However, these mechanisms target discrete GPUs. In edge unified memory architectures (UMA), such heavy-handed state management is obsolete. \sysname{} exploits UMA for in-place KV metadata freezing, achieving zero-overhead context switching without moving physical bytes.

\noindent\textbf{Dynamic Speculative Scheduling.}
Recent algorithms push acceptance rates via training-time testing (EAGLE-3~\cite{li2025eagle3}) and pipelined SSM voting (Smurfs~\cite{wang2025towards}). System-level designs tailor speculation budgets to specific constraints: AdaServe~\cite{li2026adaserve} optimizes for cloud SLOs, EdgeLLM~\cite{xu2025edgellm} adapts draft-tree width for memory-constrained devices, and Ghidorah~\cite{wei2025ghidorah} partitions verification across edge cores. For agents, DualSpec~\cite{zhong2026dualspec} utilizes semantic critics, and AgentInfer~\cite{lin2025towards} leverages suffix automata. Unlike these coarse-grained or static allocations, \sysname{} uses the HAL as an effective indicator of drafting difficulty to dynamically allocate budgets across concurrent multi-agent streams.

\noindent\textbf{Heterogeneous Execution on Edge SoCs.}
To harness disparate edge compute units, HeteroInfer~\cite{chen2025characterizing}, Doppeladler~\cite{jiang2025doppeladler}, and earlier vision frameworks~\cite{li2025accelerating} predict workload split ratios across heterogeneous processing units. Concurrently, Hello SME!~\cite{remke2024hello} and MPGEMM~\cite{deng2025demystifying} optimize standalone GEMM micro-kernels for ARM SME architectures. However, the former execution frameworks rely on per-operator runtime heuristics without whole-graph analysis, while the latter isolated libraries lack end-to-end integration. \sysname{} bridges this gap via a logical DAG topology barrier, enabling lock-free, zero-copy tensor routing to saturate asymmetric cores without runtime restructuring.

\revised{\noindent\textbf{Generalizability beyond UMA.}
EdgeAgent's two layers have different portability. The agent-aware scheduling layer is memory-topology-agnostic: HAL-based budget allocation and suspend-and-yield operate on runtime state (drafting difficulty and tool-stall events), which exists in any speculative-decoding system. They therefore transfer unchanged to discrete-GPU datacenters, where tool-induced stalls cause the same slot-wasting and head-of-line blocking problem. The UMA-aware execution layer, in contrast, derives its benefit specifically from the coherent shared address space: on a discrete GPU over PCIe the disjoint-write trick cannot remove the host–device transfer, although the same principle, partitioning by device throughput and synchronizing only on completion, maps to multi-GPU tensor parallelism over high-bandwidth interconnects or to CPU-offload schemes such as FlexGen~\cite{sheng2023flexgen}, where the goal becomes minimizing PCIe traffic and overlapping compute with transfers. Finally, the ARM SME micro-kernels are a backend accelerator rather than an architectural dependency: they speed up the CPU shard but are not required for correctness, and equivalent tiled kernels exist for x86 AMX and NVIDIA Tensor Cores.}
\section{Conclusion} 
\label{sec:conclusion}

We propose \sysname{}, a cross-layer inference system that tightly couples agent-aware scheduling with hardware-level execution to meet the demands of concurrent multi-agent LLM workflows on edge UMA devices. By leveraging agent-aware dynamic budgeting, stall-aware suspend-and-yield mechanisms, and zero-copy architecture-aware tensor parallelism, \sysname{} largely improves hardware utilization even under highly irregular execution patterns. Our design illustrates the necessity of end-to-end co-design across semantic and physical layers for efficient, real-time agentic AI, paving the way for future intelligent edge systems.

\begin{acks}
This research was supported by the National Key R\&D Program of China (2025YFB3003701, 2025YFB3003700),
the National Laboratory Special Program of Peng Cheng Laboratory under Grant No. PCL2026A04,
the National Natural Science Foundation of China under Grant No.62402534, the Yunnan Provincial Major Science and Technology Special Plan Projects under Grant No.202502AD080009, the Pazhou Lab Research Project No.PZL2023KF0001, the Guangxi Key Research and Development Program (GuikeAB25069495), and the Fundamental Research Funds for the Central Universities, Sun Yat-sen University (23xkjc016).
\end{acks}

\bibliographystyle{ACM-Reference-Format}
\balance
\bibliography{refs}

\end{document}